\documentclass[11pt,hyper,a4paper]{article}
\usepackage{jheppub}
\usepackage{amsmath,amssymb,amsfonts,amscd,bm, amsthm, slashed, mathrsfs, bm}
\usepackage{graphicx, wrapfig}
\usepackage{multirow}
\usepackage{verbatim}
\usepackage[title]{appendix}
\usepackage{fancybox}
\usepackage[utf8]{inputenc}
\usepackage{arydshln}
\usepackage{dsfont}

\usepackage{url}
\usepackage{float}

\usepackage[normalem]{ulem}
\usepackage{graphicx}

\usepackage{color}
\usepackage[usenames,dvipsnames,svgnames,table]{xcolor}
\definecolor{darkblue}{cmyk}{0.9,0.9,0,0}
\hypersetup{colorlinks=false, linkcolor=darkblue, linkcolor=darkblue, citecolor=darkblue}

\usepackage{bbm}
\usepackage{enumitem}
\usepackage{float}
\usepackage{empheq}
\usepackage{enumerate}

\title{3d-3d correspondence and abelian flat connection}

\author{Hee-Joong Chung}

\affiliation{Department of Science Education, Jeju National University, Jeju, 63243, Republic of Korea}

\abstract{We realize a homological block of a knot complement in $S^3$ for $G_{\mathbb{C}}=SL(2,\mathbb{C})$ as a half-index of a 3d $\mathcal{N}=2$ theory via an expression of the homological block as an inverted Habiro series by working out some examples, which we expect to extend to general knots.
Also, by choosing a certain set of poles in the integral expression of the half-index, we obtain the colored Jones polynomial.}

\begin{document}

\maketitle


\section{Introduction}

The 3d-3d correspondence relates 3d complex Chern-Simons theory on 3-manifold $M_3$ and 3d $\mathcal{N}=2$ Chern-Simons matter theories $T[M_3]$ \cite{Dimofte-Gaiotto-Gukov, Dimofte-Gukov-Hollands, Terashima-Yamazaki}, which helps to better understand each part of the correspondence.
In its early development, the correspondence was systematically built up by gluing ideal tetrahedra \cite{Dimofte-Gaiotto-Gukov}.
Each ideal tetrahedron corresponds to a chiral multiplet with background Chern-Simons terms in 3d $\mathcal{N}=2$ abelian theories and the gluing operation is related to the $SL(2,\mathbb{Z})$ action on 3d $\mathcal{N}=2$ abelian theories.
This is applied to knot complements in $S^3$ except the case of the unknot.

Though this construction contains non-abelian branches or flat connections, it didn't include abelian flat connections, so couldn't produce a partition function that contains all flat connections such as Jones polynomials.
This was due to the fact that ideal tetrahedron itself doesn't have the abelian flat connection.
Accordingly, it was expected that the 3d $\mathcal{N}=2$ theories constructed in \cite{Dimofte-Gaiotto-Gukov} wouldn't be a full theory since they didn't capture all the flat connections, in particular the abelian flat connections as emphasized in \cite{Chung-Dimofte-Gukov-Sulkowski}.
There have been some efforts to find or construct a full theory $T_{\text{full}}[M_3]$ that captures all flat connections including the abelian flat connections, but so far such a construction has not been available.	\\

The homological block was introduced in \cite{Gukov-Putrov-Vafa, Gukov-Pei-Putrov-Vafa, Gukov-Manolescu} a few years after the introduction of the 3d-3d correspondence.
Unlike the partition function that was constructed in \cite{Dimofte-Gaiotto-Gukov}, the homological block is regarded as the contribution from the abelian flat connection to the analytically continued Chern-Simons partition function.
Interestingly, it also knows the contribution from the non-abelian flat connections, and this can be seen from resurgent analysis \cite{Gukov-Marino-Putrov, Chung-resurg}.
Therefore, it is expected that the homological block is a natural quantity that arises from an appropriate $T[M_3]$ in the 3d-3d correspondence, where it corresponds to a half-index of a 3d $\mathcal{N}=2$ theory $T[M_3]$ with 2d $\mathcal{N}=(0,2)$ boundary conditions.

Though there are formulas to calculate homological blocks for a number of 3-manifolds, it was still not available how to realize them as half-indices of 3d $\mathcal{N}=2$ theories, in particular in the case of knot complements, when the gauge group is $G=SU(2)$ or of a higher rank.	\\

There are two types of expansions of the homological block for a knot complement $S^3 \backslash K$ in $S^3$ \cite{Gukov-Manolescu}.
The balanced expansion is $q$, $x$-series that is invariant under the Weyl action $x \leftrightarrow x^{-1}$, so it is an appropriate quantity for $G=SU(2)$.
There is a positive (resp. negative) expansion, which is a part of the balanced expansion with positive (resp. negative) powers of $x$.
These have equivalent information in the sense that the balanced expansion can be obtained by taking the average of the positive expansion under the Weyl action.
The positive expansion admits the quiver form or the fermionic sum expression and this can be realized as a half-index of an abelian quiver 3d $\mathcal{N}=2$ theory with appropriate boundary conditions \cite{Chung-3d3dpe}.
However, the balanced expansion has not been obtained as a half-index of a 3d $\mathcal{N}=2$ theory.

Meanwhile, the balanced expansion of the homological block is conjectured to be expressed as the inverted Habiro series, in the sense that when expanded as a power series in $x$ it agrees with the positive expansion of the homological block \cite{Park-inverted}.
The inverted Habiro series provides a closed-form expression of the balanced expansion, so it would be an appropriate expression for the half-index.	\\

In this work, we realize a homological block of a knot complement $S^3 \backslash K$ for $G_{\mathbb{C}}=SL(2,\mathbb{C})$ in the form of the inverted Habiro series as a half-index of a 3d $\mathcal{N}=2$ theory with certain choices of boundary conditions and an integration contour by working out some examples, which we expect to extend to general knots.
By doing so, we obtain a 3d $\mathcal{N}=2$ theory $T[M_3]$ that knows all branches of flat connections.
We also obtain the Jones polynomial, which contains all flat connections, by choosing another set of poles.	\\

In section \ref{sec:ab}, we discuss the case of the figure-eight knot, which is the simplest example, and the trefoil knots.
Considering an example other than a knot, we see that the appearance of the abelian branch arising from an appropriate choice of contour in the half-index would be specific to the case of a knot where the abelian branch is always present.
In addition, from the study of a critical point for the abelian branch and a contour associate to it, we also see that the contour that we choose for the calculation of homological blocks would be a natural one that passes through the critical point for the abelian branch.
In section \ref{sec:conc-disc}, we formulate a general expectation for arbitrary knots and discuss some aspects of homological blocks as half-indices obtained in section \ref{sec:ab}.
In Appendix \ref{sec:appendix}, we discuss certain aspects of an additional branch of the SUSY parameter space in the case of the trefoil knots.


\section{Abelian flat connection and half-index of $T[M_3]$}
\label{sec:ab}

In the 3d-3d correspondence, the partition function of the analytically continued Chern-Simons theory on 3-manifold $M_3$ corresponds to the half-index of a 3d $\mathcal{N}=2$ theory $T[M_3]$ on $D^2 \times_q S^1$ with boundary conditions preserving 2d $\mathcal{N}=(0,2)$ supersymmetry.
We consider the case $M_3= S^3 \backslash K$, a knot complement of a knot $K$ in $S^3$.
Since the homological block is the contribution from the abelian flat connection to the analytically continued Chern-Simons partition function and also encodes that from the non-abelian flat connection, it is natural to consider the homological block as a half-index of a 3d $\mathcal{N}=2$ theory $T[M_3]$.

The homological block $F_K(x,q)$ of a knot for $G_\mathbb{C}=SL(2,\mathbb{C})$ was originally obtained as a $q,x$-series expansion with integer powers and coefficients in \cite{Gukov-Manolescu}.
The balanced expansion is a Laurent series in $q$ with coefficients in the Laurent polynomial ring $\mathbb{Z}[x^{\pm1}]$, while the positive (resp. negative) expansion contains only positive (resp. negative) powers of $x$.
The balanced expansion of the homological block would be an appropriate quantity to consider in the 3d-3d correspondence for $G_\mathbb{C}=SL(2,\mathbb{C})$, considering that it is invariant under the Weyl action $x \leftrightarrow x^{-1}$.
It was conjectured that the homological block $F_K(x,q)$ for a knot $K$ can be expressed as an inverted Habiro series, which provides a closed-form expression, in the sense that when expanded it agrees with $F_K$ as a power series in $x$ \cite{Park-inverted}.	\\

The inverted Habiro series is obtained from the Habiro cyclotomic expansion of the colored Jones polynomial. 
The Habiro cyclotomic expansion of the $n$-colored Jones polynomial $J_K(n,q)$ for the $n$-dimensional representation of $G=SU(2)$ is given by
\begin{align}
J_K(n,q) = -\sum_{k=0}^{\infty} a_k(K;q) \prod_{j=1}^{k} (x+x^{-1}-q^{j}-q^{-j}) \bigg|_{x=q^n}
\label{hsjs}
\end{align}
where $a_k(K,q)$, $k=0, 1, \ldots$, is a Laurent polynomial in $q$ with integer coefficients \cite{Habiro-series}.
From \eqref{hsjs}, the inverted Habiro series is given by 
\begin{align}
F_K(x,q) = \sum_{k=1}^{\infty} \frac{a_{-k}(K,q)}{\prod_{j=0}^{k-1} (x+x^{-1}-q^{j}-q^{-j})}	
\label{invhhb}
\end{align}
where $k$ in $a_k(K;q)$ is extended to negative integers.
For $k<0$, $a_k(K;q)$ is, in general, a Laurent series in $q$ with integer coefficients.
The expression \eqref{invhhb} gives a normalized version of the homological block.
The unnormalized version of the homological block is obtained by multiplying \eqref{invhhb} by the homological block of the unknot, $x^{\frac{1}{2}} - x^{-\frac{1}{2}}$.


\subsection{Half-index for homological block of figure-eight knot}
\label{sec:cal1}

We first consider the figure-eight knot complement in $S^3$. 
The inverted Habiro series expression of the homological block for the figure-eight knot is given by
\begin{align}
F_{4_1}(x,q) = \sum_{m=0}^{\infty} \frac{1}{\prod_{j=0}^{m}(x+x^{-1} -q^j - q^{-j})}	\,	,
\end{align}
which is also expressed as
\begin{align}
F_{4_1}(x,q) = \sum_{k=0}^{\infty} (-1)^k q^{\frac{1}{2}k(k+1)} \frac{(q^{k+1} x;q)_\infty (q^{k+1} x^{-1} ;q)_\infty}{(x;q)_{\infty} (x^{-1};q)_{\infty}} 	\,	.
\label{hb41}
\end{align}

The normalized version of the homological block \eqref{hb41} is annihilated by the quantum $\hat{A}$-polynomial
\begin{align}
\begin{split}
&\hspace{-5mm}\big( (1 + q \hat{x}) (q \hat{x}^2 -1) \hat{y} - q (q^2 \hat{x} +1) (q^5 \hat{x}^2-1) \big)	\\
&\hspace{-3mm}\times \big( q^2 \hat{x}^2 (q^2 \hat{x}-1) (q \hat{x}^2-1) \hat{y}^2 - (q \hat{x} -1)^2 (q^5 \hat{x}^5 - q(1 + q + q^2) \hat{x}^2(q\hat{x}+1) + 1) \hat{y}	\\
&\hspace{5mm}+ q^2 (\hat{x}-1) \hat{x}^2 (q^3 \hat{x}^2-1) \big)
\end{split}
\label{qA41}
\end{align}
where $\hat{x} f(x) = x f(x)$, $\hat{y} f(x) = f(qx)$, and $\hat{y} \hat{x} = q \hat{x} \hat{y}$.
The quantum $\hat{A}$-polynomial \eqref{qA41} is obtained by using the {\it Mathematica} package {\tt HolonomicFunctions} \cite{Koutschan:holofunctions}, which is also used for other examples in this paper.	
By taking a classical limit $q \rightarrow 1$, the classical $A$-polynomial, or simply the $A$-polynomial, is obtained as a factor
\begin{align}
(x-1)^3 (x+1)^3 (y-1) (x^2 y^2 - ( x^4 - x^3 - 2 x^2 - x + 1) y +x^2)	\,	,
\label{A41}
\end{align}
where $x$ and $y$ are eigenvalues of the holonomies along the meridian and longitude cycles of a knot complement in $S^3$, respectively.
In \eqref{A41}, we see that there is an abelian branch $y-1$.	
Therefore, homological block \eqref{hb41} contains the abelian branch.	\\

The $A$-polynomial can also be obtained from a twisted superpotential. 
From the asymptotic expansion of \eqref{hb41}, the twisted superpotential is obtained as
\begin{align}
\widetilde{\mathcal{W}}_{4_1}(z,x) = \frac{1}{2} (\log z)^2 + \log (-1) \log z + \text{Li}_2(zx) + \text{Li}_2(zx^{-1}) + \frac{1}{2} ( \log(-x))^2	\,	.
\label{twsp41}
\end{align}
By using the supersymmetric vacuum condition,
\begin{align}
1= \exp \frac{\partial \widetilde{W}(z,x)}{\partial \log z}	\,	,	\qquad	y=\exp \frac{\partial \widetilde{W}(z,x)}{\partial \log x} 	\,	,
\label{susycond}
\end{align}
we obtain 
\begin{align}
y^2 - (x^2 - x^1 - 2 - x^{-1} + x^{-2}) y + 1	\,	,
\end{align} 
which is the non-abelian branch of the $A$-polynomial.
We see that the abelian branch is not obtained in this way, though the homological block does contain the abelian branch.
This is because the contribution of the abelian flat connection to the twisted superpotential vanishes if there are no deformation parameters that can keep it manifestly.\footnote{For example, such deformation parameters for global symmetries can be a $t$-parameter from homological-flavor locking, an $a$-parameter from the large $N$ limit, or both \cite{FGS-superA, Gukov-Putrov-Vafa, Chung-Dimofte-Gukov-Sulkowski}.}
For example, for the unknot complement in $S^3$, the only flat connection is abelian, and the twisted superpotential obtained from the asymptotic expansion of the homological block of the unknot vanishes.	\\

We would like to produce \eqref{hb41} as a half-index.
We consider an expression
\begin{align}
(q;q)^2_\infty \frac{1}{(x;q)_{\infty} (x^{-1};q)_{\infty}} \int \frac{dz}{2\pi i z} \frac{1}{\theta((-q^{\frac{1}{2}})z ;q)} (qzx;q)_\infty (qzx^{-1};q)_\infty
\label{hb41int}
\end{align}
where
\begin{align}
\theta(y;q) := (-q^{\frac{1}{2}} y;q)_\infty (-q^{\frac{1}{2}} y^{-1};q)_\infty
\label{theta}
\end{align}
is a Jacobi theta function divided by $(q;q)_\infty$.
The integrand of the integral arises from the contributions to the half-index from the field contents under the $U(1)_z \times U(1)_x \times U(1)_R$ symmetry, subject to boundary conditions:
a $U(1)_z$ vector multiplet with Neumann boundary condition;
four 3d chiral multiplets, two with Dirichlet boundary conditions and charges $(-1,-1,0)$ and $(-1,1,0)$, and two with Neumann boundary conditions and charges $(0,\pm1,0)$;
a 2d chiral multiplet with charges $(1,0,2)$;
and a free 3d chiral multiplet with Dirichlet boundary condition.
The anomaly is cancelled in this theory, and there is no need to introduce a background Chern-Simons term. 
Upon $z \rightarrow z^{-1}$, this expression is the same as the integral expression of the half-index for the figure-eight knot appeared in \cite{Beem-Dimofte-Pasquetti}, up to an overall factor.

The integral \eqref{hb41int} was evaluated in \cite{Beem-Dimofte-Pasquetti} when $|q|>1$ by choosing half-infinite lines of poles from 3d chiral multiplets and the resulting expressions are given by the Hahn-Exton $q$-Bessel function.
Using the properties that it is convergent both for $|q|>1$ and $|q|<1$, the half-index for $|q|<1$ was proposed with several consistency checks rather than being calculated directly along the contours that were suggested.
It was also shown that such half-indices correspond to non-abelian flat connections and they are annihilated by the quantum $\hat{A}$-polynomial for the non-abelian branch.	\\

Instead of these, we choose a different contour.
We first calculate the integral by choosing a zeroth coefficient of the integrand of \eqref{hb41int}, which is equivalent to taking the poles within a unit circle $|z|=1$.
Such a choice of contour was discussed, for example, in \cite{CCFFGHP} in the context of the 3d-3d correspondence.

From \cite{Garvan}, it is known that
\begin{align}
\frac{1}{(qz;q)_\infty (qz^{-1};q)_\infty} = \frac{1}{(q;q)_\infty^2} \sum_{m=-\infty}^{+\infty} \bigg( \sum_{k=1}^{\infty} (-1)^{k-1} q^{\frac{1}{2}k(k-1)+k|m|} (1-q^k) \bigg) z^m
\label{garvanid}
\end{align}
for $|qz|, |qz^{-1}|<1$.
In addition, when $|z|<\infty$, 
\begin{align}
(z;q)_\infty = \sum_{n=0}^{\infty} \frac{(-1)^n q^{\frac{1}{2}n(n-1)} z^n}{(q;q)_n}	\,	.
\label{pochid}
\end{align}
By using \eqref{garvanid} and \eqref{pochid}, the integral \eqref{hb41int} can be expressed as
\begin{align}
\begin{split}
\frac{1}{(x;q)_{\infty} (x^{-1};q)_{\infty}} \int \frac{dz}{2\pi i z} & \sum_{m=-\infty}^{+\infty} \bigg( \sum_{k=1}^{\infty} (-1)^{k-1} q^{\frac{1}{2}k(k-1)+k|m|} (1-q^k) \bigg) \frac{(qz)^m}{1-qz} \\
& \times \sum_{n=0}^{\infty} \frac{(-1)^n q^{\frac{1}{2}n(n-1)} (qzx)^n}{(q;q)_n} \sum_{l=0}^{\infty} \frac{(-1)^l q^{\frac{1}{2}l(l-1)} (qzx^{-1})^l}{(q;q)_l}	\,	.
\end{split}
\label{hb41intstep1}
\end{align}
Expanding $(1-qz)^{-1}$ as a geometric series, we have terms $(qz)^{m+n+l} \sum_{\alpha=0}^{\infty} (qz)^{\alpha}$ in the integrand.
The residue integral picks $m=-n-l-\alpha$, and \eqref{hb41intstep1} becomes
\begin{align}
\hspace{-0.7mm}\sum_{\alpha=0}^{\infty} \sum_{k=1}^{\infty} (-1)^{k-1} q^{\frac{1}{2}k(k-1)+k(n+l+\alpha)} (1-q^k) \sum_{n=0}^{\infty} \frac{(-1)^n q^{\frac{1}{2}n(n-1)} x^n}{(q;q)_n} \sum_{l=0}^{\infty} \frac{(-1)^l q^{\frac{1}{2}l(l-1)} x^{-l}}{(q;q)_l}
\label{hb41intstep2}
\end{align}
The sum over $\alpha$ gives $\sum_{\alpha=0}^{\infty} q^{k\alpha} = (1-q^k)^{-1}$, and this cancels the factor $(1-q^k)$ in \eqref{hb41intstep2}.
Therefore, by using \eqref{pochid}, we obtain
\begin{align}
\sum_{k=0}^{\infty} (-1)^{k} q^{\frac{1}{2}k(k+1)} \frac{(q^{k+1}x;q)_\infty (q^{k+1}x^{-1};q)_\infty}{(x;q)_{\infty} (x^{-1};q)_{\infty}} 	\,	,	
\label{hb41intstep3}
\end{align}
and this agrees with the normalized version of the homological block \eqref{hb41} for the figure-eight knot. 	\\

The homological block \eqref{hb41intstep3} can also be obtained by taking a contour as a unit circle $|z|=1$ and calculate the residues for the poles $z=q^k$, $k=0,1, \ldots$ from $(z^{-1};q)_\infty^{-1}$ of $\theta((-q^{\frac{1}{2}})z;q)^{-1}$ in \eqref{hb41int}.
More precisely, this case is special in that poles are only from $\theta((-q^{\frac{1}{2}})z;q)^{-1}$.
In general, when there are other poles, the contour that is responsible for the abelian branch is the contour that encloses poles from $(z^{-1};q)_\infty^{-1}$ as we will see.	\\

Interestingly, it is possible to obtain the colored Jones polynomial of the figure-eight knot from \eqref{hb41int} by choosing another set of poles.
Instead of choosing poles $z=q^k$, $k=0,1, \ldots$, if we take the poles $z=q^{-k}$, $k=1, 2, \ldots, n$ from $(qz;q)_\infty^{-1}$ of $\theta((-q^{\frac{1}{2}})z;q)^{-1}$ in \eqref{hb41int} we have
\begin{align}
-\sum_{k=0}^{n-1} (-1)^k q^{-\frac{1}{2}k(k+1)} (q x^{-1};q)_{k} (q x;q)_{k}	\,	.
\end{align}
When $x=q^n$ where $n$ denotes the $n$-dimensional representation of $SU(2)$, this yields
\begin{align}
J_{4_1}(n,q) = -\sum_{k=0}^{n-1} (-1)^k q^{-\frac{1}{2}k(k+1)} (q^{1-n};q)_{k} (q^{1+n};q)_{k}	\,	,
\end{align}
which agrees with the expression of the $n$-colored Jones polynomial of the figure-eight knot in \cite{Garoufalidis-Le}.	\\

We can also consider another expression corresponding to different boundary conditions, 
\begin{align}
(q;q)_\infty^2 (qx;q)_\infty (qx^{-1};q)_\infty \int \frac{dz}{2 \pi i z} \frac{1}{\theta((-q^{\frac{1}{2}})z;q) \, (z^{-1}x;q)_\infty (z^{-1} x^{-1};q)_\infty} \frac{\theta(z;q) \theta(qz;q)}{\theta(1;q) \theta(q;q)}	\,	,
\label{hb41int2}
\end{align}
which also gives the homological block and the Jones polynomial from the same sets of poles.
Compared to the previous case \eqref{hb41int}, Dirichlet and Neumann boundary conditions of the four 3d chiral multiplets are interchanged.
Also, additional 2d $\mathcal{N}=(0,2)$ chiral and Fermi multiplets are introduced: chiral multiplets with charges $(0,0,1)$ and $(0,0,3)$, and Fermi multiplets with charges $(1,0,0)$ and $(1,0,2)$.

In the integral \eqref{hb41int2}, in addition to the familiar poles from $\theta((-q^{\frac{1}{2}})z;q)^{-1}$, there are other poles from $(z^{-1}x;q)_\infty^{-1}$ and $(z^{-1} x^{-1};q)_\infty^{-1}$, which capture non-abelian branches explicitly.
Indeed, by taking poles $z=x^{\pm 1}q^k$ from $(z^{\mp}x;q)_\infty^{-1}$, \eqref{hb41int2} becomes, respectively,
\begin{align}
- \frac{1}{1-x^{\pm 1}} \frac{1}{(q;q)_\infty} \left( \frac{\theta(x;q)}{\theta(1;q)} \right)^2 \frac{1}{\theta((-q^{\frac{1}{2}})x^{\pm2};q)} J(x^{\pm1},x^{\pm2};q)
\label{hb41nab}
\end{align}
where $J(x,y;q)$ is the Hahn-Exton $q$-Bessel function
\begin{align}
J(x,y;q) = (qy;q)_\infty \sum_{k=0}^{\infty} \frac{(-1)^{k} q^{\frac{1}{2}k(k+1)} x^k}{(q;q)_k (qy;q)_k}	\,	.
\end{align}
The half-indices \eqref{hb41nab} are annihilated by 
\begin{align}
\begin{split}
&q^2 \hat{x}^2 (q^2 \hat{x} -1) (q \hat{x}^2-1) \hat{y}^2 
-(q \hat{x}-1)^2 \big(q^5 \hat{x}^5-q (1+q+q^2) \hat{x}^2 (q \hat{x}+1)+ 1 \big) \hat{y}	\\
&\hspace{90mm}+q^2 (\hat{x}-1) \hat{x}^2 (q^3 \hat{x}^2 -1)	\,	.
\end{split}
\end{align}
Upon the classical limit, it becomes
\begin{align}
-(x-1)^2 (x+1) x^2 \big( y^2 - (x^2-x-2-x^{-1}+x^{-2} ) y+1\big)	\,	,
\end{align}
which contains the $A$-polynomial for non-abelian branch as a factor.
Thus, this example shows that the contour enclosing poles from $(z^{-1};q)_\infty^{-1}$ is associated to the abelian branch, while the contours enclosing poles from $(z^{\mp}x;q)_\infty^{-1}$ are associated to the non-abelian branch.
Therefore, the $T[S^3 \backslash 4_1]$ described above for \eqref{hb41int2} explicitly contains all branches of flat connections.


\subsection{Trefoil knots}
\label{sec:trefoil}

We also consider the cases of the left-handed and right-handed trefoil knots.
The expressions of the normalized homological block as the inverted Habiro series for the left-handed and right-handed trefoil knots are given \cite{Park-inverted}, respectively, by
\begin{align}
F_{3_1^l}(x,q) &= -q^{-1} \sum_{k=0} \frac{q^{k^2}}{(x;q)_{k+1} (x^{-1};q)_{k+1}}		\,	,	\label{f31l}	\\
F_{3_1^r}(x,q) &= -q \sum_{k=0} \frac{q^{k}}{(x;q)_{k+1} (x^{-1};q)_{k+1}}	\,	.	\label{f31r}
\end{align}
When expanded, they agree with the known results for the $q,x$-series expansion of the homological blocks.
In the case of the figure-eight knot, it is possible to have an integral expression such that poles are only from the Jacobi theta function in the denominator.
In the case of trefoil knots, the expressions \eqref{f31l} and \eqref{f31r} lead to integral expressions that contain other poles.	
Since the cases of the left-handed and the right-handed trefoil knots exhibit some differences, we discuss them separately.


\subsubsection*{Left-handed trefoil knot}

The homological block \eqref{f31l} for the left-handed trefoil knot is annihilated by 
\begin{align}
\big( (q \hat{x}^2-1) \hat{y} + q(1-q^3 \hat{x}^2) \big)
\big( (q \hat{x}-1) \hat{y} +  q^2 \hat{x}^3 (\hat{x} - 1) \big)	\,	.
\label{qa31l}
\end{align}
When taking the classical limit, \eqref{qa31l} becomes
\begin{align}
(x-1)^2 (x+1) (y-1) (y + x^3)	\,	,
\label{31lfull}
\end{align}
and it contains the abelian branch $y-1$ as a factor.

From the twisted superpotential 
\begin{align}
\widetilde{\mathcal{W}}_{3_1^l} &= ( \log z )^2 + \frac{1}{2} \big( \log(-x) \big)^2 + \text{Li}_2(zx) + \text{Li}_2(zx^{-1})	\,	,
\label{twsp31l}
\end{align}
the SUSY parameter space is obtained,\footnote{\label{31lnbr}
More precisely, with the twisted superpotential \eqref{twsp31l}, there is another solution $z =\infty$ in the critical point equation $1= \exp \partial \widetilde{\mathcal{W}}_{3_1^l} / \partial \log z$, which leads to $xy+1=0$ from $y= \exp \partial \widetilde{\mathcal{W}}_{3_1^l} / \partial \log x$.
We discuss more about this in Appendix \ref{sec:appendix}.}
\begin{align}
y+ x^3 = 0		\,	.
\label{31lnab}
\end{align}
It is the $A$-polynomial of the non-abelian branch, and the abelian branch $y-1$ is not obtained at the level of the twisted superpotential as in the case of the figure-eight knot.	\\

We consider the integral
\begin{align}
(q;q)_\infty^2 \int \frac{dz}{2 \pi i z} \frac{1}{\theta(-q^{\frac{1}{2}} z ;q)} \frac{(qzx;q)_\infty}{(z^{-1}x;q)_\infty} \frac{\theta((-q^{\frac{1}{2}})^3 z x^{-3} ;q)}{\theta((-q^{\frac{1}{2}})x^{-3};q)} \frac{\theta((-q^{\frac{1}{2}}) x^2 ;q)}{\theta( (-q^{\frac{1}{2}}) z x^{-2};q)}	\,	.
\label{int31l}
\end{align}
The integrand is obtained from the contribution to the half-index from the following field contents under the $U(1)_z \times U(1)_x \times U(1)_R$ symmetry, with boundary conditions: 
a $U(1)_z$ vector multiplet with Neumann boundary condition; 
two 3d chiral multiplets with Dirichlet and Neumann boundary condition and charges $(-1,-1,0)$ and $(-1,1,0)$, respectively; 
three 2d chiral multiplets with charges $(1,0,2)$, $(0,-3,2)$, and $(1,-2,2)$; 
a 2d Fermi multiplet with charges $(1,-3,3)$ and $(0,2,1)$; 
and a free 3d chiral multiplet with Dirichlet boundary condition.
There is a gauge-invariant anti-monopole operator with charges $(0,0,4)$.
The UV Chern-Simons terms that are encoded in the anomaly polynomial $\mathbf{f}_z (\mathbf{f}_z - 4\mathbf{f}_R) + 4 \mathbf{f}_x \mathbf{f}_R- \frac{15}{2}\mathbf{f}_R^2$ are added to cancel the anomaly where $\mathbf{f}_*$ denotes a field strength of $U(1)_*$.	\\

This integral has the poles from $\theta(-q^{\frac{1}{2}} z ;q)^{-1}$, $(z^{-1}x;q)_\infty^{-1}$, and $\theta( (-q^{\frac{1}{2}}) z x^{-2};q)^{-1}$.
Unlike the case of the figure-eight knot, there is no integral expression that has poles only from $\theta(-q^{\frac{1}{2}} z ;q)^{-1}$ for \eqref{f31l}.
Therefore, we take the factors $\theta(-q^{\frac{1}{2}} z ;q)^{-1} (qzx;q)_\infty  (z^{-1}x;q)_\infty^{-1}$ in the integrand, and in this setup an integral expression for \eqref{f31l} is not available without at least an extra $z$-dependent theta function in the denominator.
This leads to extra poles from the $z$-dependent theta function such as $\theta( (-q^{\frac{1}{2}}) z x^{-2};q)^{-1}$ in \eqref{int31l}, but the detail can vary as there is an ambiguity in choosing boundary degrees of freedom.
The quantity obtained by taking a set of poles $z= x^2 q^k$, $k=0,1, \ldots$ is annihilated by the operator\footnote{This operator is given by $\big( \, \big( (q \hat{x}-1)^2  (q^4 \hat{x}^3-1) (q^5 \hat{x}^3-1) (q^2 \hat{x}^2+q \hat{x}+1) (q \hat{x}^3 (\hat{x}-1) (q^2 \hat{x}^3 - q\hat{x} -q -1)-1 ) \big) \hat{y} 
+ q^8 \hat{x}^9 \big( q^4 \hat{x}^3 (q \hat{x}-1) (q^5 \hat{x}^3 - q^2 \hat{x} - q-1)-1 \big) \, \big)
( \hat{y} + q^2 \hat{x}^3 )$.} that contains the usual quantum $\hat{A}$-polynomial for the non-abelian branch and additional factors that depend on the choice of theta functions in the integrand.
In this case, there are two critical points corresponding to non-abelian and abelian branches, to each of which we associate an integration contour that passes through it.
Since there is no further critical point for $z \neq 0, \infty$, there wouldn't be a further contour to encloses the poles from the extra theta function.\footnote{Since $z \in \mathbb{C}^*$, it would not be appropriate to associate the contour to $z=0$ and $\infty$.}
Therefore, in this case, we expect that it is not appropriate to have as a half-index the quantity obtained by evaluating the contour integral that encloses poles from such an additional theta function.
Thus, we only consider the poles from $\theta(-q^{\frac{1}{2}} z ;q)^{-1}$ and $(z^{-1}x;q)_\infty^{-1}$.	\\

When taking poles $z=q^k$, $k=0,1, \ldots$ from $(z^{-1};q)_\infty^{-1}$ of $\theta(-q^{\frac{1}{2}} z ;q)^{-1}$, we obtain
\begin{align}
-q^{-1} (1-x) &\sum_{k=0}^{\infty} \frac{q^{k^2}}{(x;q)_{k+1} (x^{-1};q)_{k+1}}	\,	,
\label{31labcal}
\end{align}
which is the homological block \eqref{f31l} up to overall factors.\footnote{The factor $(1-x)$ can be removed if we include the term $\frac{(qx;q)_\infty}{(x;q)_\infty}$ in \eqref{int31l}.}

If we choose the poles $z=xq^k$, $k=0,1, \ldots$, from $(z^{-1}x;q)_\infty^{-1}$, we obtain
\begin{align}
\frac{(q;q)_\infty \theta((-q^{\frac{1}{2}})^3 x^{-2};q) \theta((-q^{\frac{1}{2}}) x^{2};q)}{\theta((-q^{\frac{1}{2}})x;q) \theta((-q^{\frac{1}{2}})x^{-1};q) \theta((-q^{\frac{1}{2}})x^{-3};q) }
\sum_{k=0}^{\infty} \frac{ q^{k^2} x^{2 k} }{(q;q)_k} (q^{k+1} x^2;q)_{\infty}	\,	,
\end{align}
which is simplified to
\begin{align}
\frac{(q;q)_\infty \theta((-q^{\frac{1}{2}})^3 x^{-2};q) \theta((-q^{\frac{1}{2}}) x^{2};q)}{\theta((-q^{\frac{1}{2}})x;q) \theta((-q^{\frac{1}{2}})x^{-1};q) \theta((-q^{\frac{1}{2}})x^{-3};q) }	\,	.
\label{31lnabcal}
\end{align}
It is annihilated by
\begin{align}
\hat{y} + q^2 \hat{x}^3	\,	,
\label{qana31l}
\end{align}
and this corresponds to the non-abelian branch \eqref{31lnab}.

We also consider taking the poles $z=q^{-k}$, $k=1,2, \ldots$ from $(qz;q)_\infty^{-1}$ of $\theta((-q^{\frac{1}{2}})z;q)^{-1}$, and it gives
\begin{align}
-(1-x) \sum_{k=0}^{\infty} q^{k} (qx;q)_k (qx^{-1};q)_k	\,	.
\label{31ljones}
\end{align}
When $x=q^n$, we obtain the Jones polynomial of the left-handed trefoil knot up to the overall factor $-(1-x)$.\footnote{Before taking the specialization $x=q^n$, \eqref{31ljones} is annihilated by
$
\big( q \hat{x} (q \hat{x}+1) (q \hat{x}^2-1) \hat{y}^2 - (q \hat{x}^2-1) (q^5 \hat{x}^2-1) \hat{y} - q (q^2 \hat{x}+1) (q^5 \hat{x}^2-1 ) \big)	
(\hat{y}+q^2 \hat{x}^3)
$.
Its classical limit is 
\begin{align}
(x-1) (x+1)^2 (y-1) (x y+1) (y+x^3)
\label{31ljp}
\end{align} 
and it contains the extra factor $xy+1$, which arises from the solution $z = \infty$ of the critical point equation $1= \exp \partial \widetilde{\mathcal{W}}_{3_1^l} / \partial \log z$.
This factor doesn't arise in the recursion relation for the colored Jones polynomial, \textit{i.e.} when $x=q^n$.
We discuss this further in Appendix \ref{sec:appendix}.
}
It is also possible to obtain the colored Jones polynomial by taking $x=q^n$ first in \eqref{int31l} then taking the poles $z=q^{-k}$, $k=1,2, \ldots, n$ in the integrand.


\subsubsection*{Right-handed trefoil knot}

The homological block for the right-handed trefoil knot \eqref{f31r} is annihilated by
\begin{align}
\begin{split}
&\big( (q \hat{x}+1) (q \hat{x}^2-1) \hat{y} + q^2 \hat{x} (q^2 \hat{x}+1) (q^5 \hat{x}^2-1) \big)
\big( (q \hat{x}^2-1) \hat{y} - q(q^3 \hat{x}^2 - 1) \big)	\\
&\hspace{85mm}\times \big( q (q \hat{x}-1) \hat{x}^3 \hat{y} + (\hat{x}-1) \big)	\,	.
\end{split}
\label{qa31r}
\end{align}
When taking the classical limit, \eqref{qa31r} becomes
\begin{align}
(x-1)^3 (x+1)^3 (y-1) (y+x) (x^3 y+1)	\,	,
\label{31rfull}
\end{align}
and it contains the abelian branch $y-1$.

From the homological block \eqref{f31r}, the twisted superpotential is obtained as
\begin{align}
\widetilde{\mathcal{W}}_{3_1^r} &= \frac{1}{2} \big(\log(-x) \big)^2 + \text{Li}_2(zx) + \text{Li}_2(zx^{-1})	\,	,
\label{twsp31r}
\end{align}
which gives the SUSY parameter space 
\begin{align}
(y+x) (y+ x^{-3}) = 0	\label{31rnab}	\,	.
\end{align}
This contains the non-abelian branch $y+ x^{-3}$ of the right-handed trefoil knot.
In addition to that, there is an additional factor $y+x$, which also appears in \eqref{31rfull}. 
By substituting two solutions $z=x+x^{-1}$ and $z=0$ of the critical point equation $1= \exp \partial \widetilde{\mathcal{W}} / \partial \log z$ into $y = \exp \widetilde{\mathcal{W}} / \partial \log x$, we have $y+x^{-3}=0$ and $y+x=0$, respectively.
Though the branch $y+x$ doesn't arise in the colored Jones polynomial, it arises also in an infinite sum expression of the positive expansion obtained from \eqref{f31r}. 
This is related to the asymptotic behavior of \eqref{f31r}, which we discuss in more detail in Appendix \ref{sec:appendix}.	\\

We consider the integral 
\begin{align}
(q;q)_\infty^2 \int \frac{dz}{2 \pi i z} \frac{1}{\theta(-q^{\frac{1}{2}} z ;q)} \frac{(qzx;q)_\infty}{(z^{-1}x;q)_\infty} \frac{\theta(z^{-1} x ;q)}{\theta(x;q)} \frac{\theta(z ;q)}{\theta(1;q)}	\,	.
\label{int31r}
\end{align}
The integrand is given by the contributions to the half-index from the following field contents under the $U(1)_z \times U(1)_x \times U(1)_R$ symmetry, with boundary conditions:
a $U(1)$ vector multiplet in Neumann boundary condition;
two 3d chiral multiplets with Dirichlet and Neumann boundary conditions and charges $(-1,-1,0)$ and $(-1,1,0)$, respectively;
three 2d chiral multiplets with charges $(1,0,2)$, $(0,1,1)$, and $(0,0,1)$;
two 2d Fermi multiplets with charges $(-1,1,0)$ and $(1,0,0)$;
and a free 3d chiral multiplet with Neumann boundary condition.
There is a gauge-invariant monopole operator with charges $(0,0,2)$.
Also, the UV Chern-Simons terms encoded in the anomaly polynomial $-\mathbf{f}_z^2 + 2\mathbf{f}_R(\mathbf{f}_z - \mathbf{f}_x)$ are added to cancel the anomaly.
In the integrand of \eqref{int31r}, there are also poles from $(z^{-1}x;q)_\infty^{-1}$ in addition to those from $\theta(-q^{\frac{1}{2}} z ;q)^{-1}$.	\\

When taking poles $z=q^k$, $k=0,1, \ldots$, from $(z^{-1};q)_\infty^{-1}$ of $\theta(-q^{\frac{1}{2}} z ;q)^{-1}$, we obtain
\begin{align}
(1-x^{-1}) &\sum_{k=0}^{\infty} \frac{q^k}{(x;q)_{k+1} (x^{-1};q)_{k+1}}	\,	,
\label{31rabcal}
\end{align}
which agree with the homological block \eqref{f31r} up to an overall factor.

When taking the poles $z=xq^k$, $k=0, 1, \ldots$ from $(z^{-1}x;q)_\infty^{-1}$, the integral becomes
\begin{align}
\frac{(q;q)_\infty}{\theta((-q^{\frac{1}{2}})x;q)} \sum_{k=0}^{\infty} \frac{q^k}{(q;q)_k} (q^{k+1} x^2;q)_\infty	\,	,
\label{int31r2}
\end{align}
and it is annihilated by
\begin{align}
\big( ( q \hat{x}^2-1 ) \hat{y} + q \hat{x}(q^3\hat{x}^2-1) \big) ( q^2 \hat{x}^3 \hat{y} + 1 )	\,	.
\end{align}
When taking a classical limit, it becomes
\begin{align}
(x^2-1) (y+x) (x^3 y + 1)	\,	,
\end{align}
so we recover \eqref{31rnab} as factors.

We can also consider taking the poles $z=q^{-k}$, $k=1,2, \ldots, n$ from $(qz;q)_\infty^{-1}$ of $\theta((-q^{\frac{1}{2}})z;q)^{-1}$ when $x=q^n$, which gives
\begin{align}
(1-x^{-1}) \sum_{k=0}^{n-1} q^{-(k+1)^2} (qx;q)_k (qx^{-1};q)_k	\bigg|_{x=q^n}	\,	.
\end{align}
Up to the overall factor $(1-x^{-1})q^{-1}$, it is the $n$-colored Jones polynomial of the right-handed trefoil knot.


\subsection{Example other than a knot}

We consider another example that is not a knot complement, and would like to see the difference.

A free chiral theory or the tetrahedron theory $\mathcal{T}$ is given by a free 3d $\mathcal{N}=2$ chiral multiplet with charge $+1$ and 0 under $U(1)_x$ global symmetry and $U(1)_R$ R-symmetry, respectively, and the UV background Chern-Simons term for $U(1)_x$ with level $-\frac{1}{2}$.
It is known that this theory forms a mirror triality with two other theories, say $\mathcal{T}'$ and $\mathcal{T}''$.

The $\mathcal{T}''$ theory is a $U(1)$ gauge theory with a chiral multiplet charged $+1$ and $0$ under the $U(1)$ gauge symmetry and the $U(1)_R$ symmetry.
In addition, there are background Chern-Simons terms that are encoded from the anomaly polynomial
$
\mathbf{I}^{\mathcal{T}''} = -\frac{1}{2} (\mathbf{f} - \mathbf{f}_R)^2 -2 (\mathbf{f} - \mathbf{f}_R) \mathbf{f}_x -\mathbf{f}_x^2 - \frac{1}{2} \mathbf{f}_R^2
$.
For Neumann boundary condition on the vector multiplet and the chiral multiplet with an additional boundary Fermi multiplet charged $+1$, $+1$, and $-1$ under the $U(1)$, $U(1)_x$, and $U(1)_R$ symmetries, respectively, the half-index is given by
\begin{align}
Z^{\mathcal{T}''}_{\mathcal{N}, N + \text{fermi}} = (q x^{-1};q)_\infty	\,	. 
\label{hindtet}
\end{align}
This half-index of $\mathcal{T}''$ agrees with that of a free chiral multiplet theory $\mathcal{T}$ with background Chern-Simons terms encoded in the anomaly polynomial $-\frac{1}{2}(\mathbf{f} - \mathbf{f}_R)^2$ \cite{Dimofte-Gaiotto-Paquette}.
It is annihilated by $\hat{x} \hat{y}-\hat{x}+1$, and there is no abelian branch, since it doesn't arise in an ideal tetrahedron.	\\

We may consider the anti-holomorphic version of the half-index, or anti-half-index, for the theory $\mathcal{T}''$ in the sense that the fusion with the half-index gives the $S^2 \times_q S^1$ index.
From \eqref{hindtet}, the anti-half-index is given by $\tilde{Z}^{\mathcal{T}''}_{1} = (x^{-1};q)_\infty^{-1}$.\footnote{If using the notation in \cite{Beem-Dimofte-Pasquetti}, the notation $q$ and $x$ in $\tilde{Z}^{\mathcal{T}''}_{1} = (x^{-1};q)_\infty^{-1}$ should be replaced by $\tilde{q}^{-1}$ and $\tilde{x}$, respectively, and these are, for example, $\tilde{q}^{-1}=q$ and $\tilde{x}=\zeta_x^{-1} q^{\frac{m_x}{2}}$ for the $S^2 \times_q S^1$ index where $\zeta_x$ and $m_x$ are a fugacity and a magnetic flux of $U(1)_x$, respectively, while $x=\zeta_x q^{\frac{m_x}{2}}$. 
For simplicity of notation, we use the variable $x$.}
It is annihilated by $(q \hat{x} -1) \hat{y} - q \hat{x}$.
Also, using the prescription in \cite{Beem-Dimofte-Pasquetti}, the integral that produces this anti-half-index would be given by
\begin{align}
\int \frac{dz}{2 \pi i z} \frac{(q z;q)_\infty}{\theta((-q^{\frac{1}{2}})xz;q)}
\label{tppint}
\end{align}
with a certain choice of a contour, which we don't specify.

Since there is one critical point in the ideal tetrahedron and no additional critical point, such as that associated to the abelian flat connection, a natural contour would be the one that gives $\tilde{Z}^{\mathcal{T}''}_{1} = (x^{-1};q)_\infty^{-1}$.
Even so, if the contour enclosing the poles $z=x^{-1} q^k$, $k=0,1, \ldots$ from $(z^{-1} x^{-1};q)_\infty^{-1}$ of $\theta((-q^{\frac{1}{2}})xz;q)^{-1}$ in \eqref{tppint} is simply chosen, we have
\begin{align}
\tilde{Z}^{\mathcal{T}''}_{2} = \frac{1}{(q;q)_\infty^2} \sum_{k=0}^{\infty} (-1)^k q^{\frac{1}{2}k(k+1)} (q^{k+1} x^{-1};q)_\infty	\,	,
\label{ahitdp}
\end{align}
and it satisfies the inhomogeneous equation $\big( (q \hat{x} -1) \hat{y} - q \hat{x} \big) \tilde{Z}^{\mathcal{T}''}_{2} = (x^{-1};q)_\infty$.
The inhomogeneous term $(x^{-1};q)_\infty$ is annihilated by $q \hat{x} \hat{y} - q \hat{x} +1$, so \eqref{ahitdp} is annihilated by $( q \hat{x} \hat{y} - q \hat{x} +1 ) \big( (q \hat{x} -1) \hat{y} - q \hat{x} \big)$.
The additional annihilator $q \hat{x} \hat{y} - q \hat{x} +1$ doesn't contain the abelian branch.
Thus, even in this case, we see that the abelian branch doesn't arise.
We expect that the appearance of the abelian branch from choosing a contour as described above would be specific to the case of a knot where the abelian branch always exists.


\subsection{Critical point and contour}
\label{ssec:contour}

Given the boundary conditions, the contours for the half-indices are associated to the critical points obtained as solutions of $1= \exp \partial \widetilde{\mathcal{W}} / \partial \log z$.
However, the critical point for the abelian branch do not arise in the twisted superpotential, which leads to the $A$-polynomial without the abelian branch.
Meanwhile, when a deformation is turned on, it is possible to keep the abelian branch manifestly, that is, the deformed $A$-polynomial contains the abelian branch explicitly.

By taking a limit that the deformation parameter is turned off, it is possible to track the critical point for the abelian branch \cite{Chung-Dimofte-Gukov-Sulkowski}.
Upon identifying a critical point for the abelian branch, we would like to see that the contour that we chose in the calculation of the homological block would be a natural convergent contour that passes through the critical point for the abelian branch.	\\

As an example, we consider the case of a trefoil knot. 
In this case, it is possible to turn on a fugacity for a global symmetry via the homological-flavor locking \cite{AS-refinedCS, Gukov-Putrov-Vafa}, and the twisted superpotential is given by
\begin{align}
\widetilde{\mathcal{W}}_{3_1^l}(z,x,t) = &\text{Li}_2((-t)^3) - \text{Li}_2(x(-t)^{3}) + \text{Li}_2(-t) - \text{Li}_2(z (-t)) + \text{Li}_2(z) + \text{Li}_2(xz^{-1})	\nonumber\\
&+ \log x \log z + 2 \log (-t) \log z - \frac{3}{2} \log (-t) \log x	\,	,
\end{align}
which can be obtained from the refined colored Jones polynomial \cite{FGS-superA, FGS-VC}.
From the equation for critical points $1= \exp \partial \widetilde{\mathcal{W}} / \partial \log z$, we have
\begin{align}
\frac{(-t)^2 x (1- z(-t)) (1-z^{-1}x)}{1-z} = 1	\,	.
\label{crit31}
\end{align}
When two solutions for $z$ are substituted into
\begin{align}
y = \exp \frac{\partial \widetilde{\mathcal{W}}}{\partial \log x} = \frac{z (-t)^{-3/2} (1 - (-t)^3 x)}{1-xz^{-1}}	\,	,
\label{susyp31}
\end{align}
the refined $A$-polynomial is obtained up to an overall factor
\begin{align}
A_{3_1^l}(x,y,t) = 
(-t)^2 (x-1) y^2
+ (-t)^{\frac{1}{2}} \left(t^6 x^4+t^5 x^3+ 2 t^2 x^2(t+1)-t^2 x+1\right) y
+(-t)^3 x^3 (t^3 x+1)	\,	.
\end{align}
By taking an unrefined limit $t \rightarrow -1$, it becomes $(x-1) (y-1) (y+x^3)$, and the unrefined limit of the refined $A$-polynomial contains the abelian branch as a factor.

When the unrefined limit $t \rightarrow -1$ is taken, one of the solutions in \eqref{crit31} becomes $z \rightarrow 1$ and $(1-z)$ factor in the numerator and the denominator of \eqref{crit31} cancels out.
While it disappears in the unrefined limit,  from \eqref{susyp31} $z=1$ leads to the abelian branch $y-1$,
\begin{align}
y =\exp \partial \widetilde{\mathcal{W}} / \partial \log x \big|_{-t=1, \, z=1} = 1	\,	.
\label{abz1}
\end{align}
The other solution gives the $A$-polynomial for the non-abelian branch.
This behavior can also be seen in the cases of the right-handed trefoil and the figure-eight knot.\footnote{More precisely, homological-flavor locking arises when $M_3$ admits a semi-free $U(1)$ action \cite{AS-refinedCS, Witten-M5knots, Gukov-Putrov-Vafa}. 
For the knot complement of the figure-eight knot in $S^3$ such a $U(1)$ action is not available, so it is not possible to turn on the refined parameter $-t$ while preserving supersymmetry. 
In such a case, we may simply deform the $A$-polynomial and the twisted superpotential to see the behavior upon the unrefined limit.
}
The condition \eqref{abz1} also arises in the examples above with \eqref{twsp41}, \eqref{twsp31l}, and \eqref{twsp31r}, all of which give
\begin{align}
y = -x \frac{1-zx^{-1}}{1-zx} \bigg|_{z=1} = 1	\,	.
\label{aby1}
\end{align}
Therefore, the contour associated to the abelian branch would pass through the critical point $z=1$.	\\

In order to make it compatible with the calculations in section \ref{sec:ab}, we shift the parameter $z$ to $(-q^{\frac{1}{2}})^{-1}z$ and take a contour that encloses poles $z=(-q^{\frac{1}{2}}) q^k$, $k=0,1, \ldots$ from $(-q^{\frac{1}{2}} z^{-1};q)_\infty^{-1}$ of $\theta(z;q)^{-1}$.
This gives the same homological block, while the critical point for the abelian flat connection is at $z=-1$.
In the $\log z$-plane, which we simply choose for clarity, this critical point is at $\log z = \pi i$, and a natural convergent contour that passes through it would be the one that encloses the poles at $\log z=\pi i + \big( k + \frac{1}{2} \big) \hbar$, $k=0,1, \ldots$, which is the contour we chose to produce the homological block.\footnote{Or we can also take the contour that extends along the imaginary axis of the $\log z$-plane, which is a cylinder, \textit{i.e.} along $|z|=1$, possibly with a deformation. 
For the examples in section \ref{sec:ab}, it captures all branches, including the abelian branch.} 
Thus, this would explain why such a choice of the contour gives rise to the homological block, since it would be a natural convergent contour that passes through the critical point for the abelian branch.


\section{Conclusions and discussion}
\label{sec:conc-disc}

In the examples discussed above, the homological block is obtained by taking the poles $z=q^k$, $k=0,1, \ldots$ from $(z^{-1};q)_\infty^{-1}$.
The usual evaluation with the contour that encloses other half-infinite lines of poles from $q$-Pochhammer symbols as in \cite{Beem-Dimofte-Pasquetti} gives a half-index annihilated by the quantum $\hat{A}_{\text{nab}}$-polynomial that doesn't contain the abelian branch. 
That is, the classical limit of such a quantum $\hat{A}$-polynomial gives the $A$-polynomial obtained from the twisted superpotential of the 3d $\mathcal{N}=2$ theory, and this doesn't include the abelian branch.
The choice of poles or the contour as in section \ref{sec:ab} leads to an inhomogeneous difference equation $\hat{A}_{\text{nab}} Z \neq 0$.
In order to make it homogeneous difference equation, additional operators or annihilators should be applied.\footnote{
The homogeneous difference equation has a physical meaning \cite{Dimofte-Gaiotto-Gukov,Dimofte-Gaiotto-Gukov-index, Beem-Dimofte-Pasquetti} as an algebra of line operators on $S^1$ acting at the tip of $D^2$ when the theory is coupled to abelian 4d $\mathcal{N}=2$ theories as a boundary theory.
An inhomogeneous difference equation by itself would not have a direct physical interpretation.
However, once it becomes a homogeneous difference equation upon applying extra annihilators, it would acquire physical meaning.
It would be interesting to understand the physical reason why choosing poles from the $(z^{-1};q)_\infty^{-1}$ factor, which together with $(qz;q)_\infty$ forms the inverse of the Jacobi theta function $\theta((-q^{\frac{1}{2}})z;q)^{-1}$ arising from the 2d boundary chiral multiplet, leads to the inhomogeneous difference equation.
}
We see that such operators contain the abelian branch for the case of a knot.
This would imply that the contour described in section \ref{sec:ab} captures the abelian branch. 
Also, the analysis on the critical point for the abelian branch would indicate that the contour that we chose to obtain the homological block is the contour associated to the abelian branch, \textit{i.e.} the contour that passes through the critical point for the abelian branch.

Put slightly differently, suppose that there is a theory with certain field contents whose SUSY parameter space obtained from the twisted superpotential by using \eqref{susycond} captures the abelian branch explicitly, which would give the full $A$-polynomial, and also that there is no deformation parameter arising from a global symmetry of the theory that can keep the abelian branch manifestly.
Then the SUSY parameter space calculated from the twisted superpotential obtained directly from the resulting half-index, \textit{i.e.} the homological block such as \eqref{hb41}, after the integration would contain the abelian branch explicitly.
However, it doesn't as we saw above.
Therefore, this perspective and the calculations in the previous section would imply that given proper boundary conditions the abelian branch of a knot complement is captured by a choice of contour rather than field contents themselves if its twisted superpotential doesn't capture the abelian branch explicitly.	\\

Also, if we choose a contour enclosing the set of poles at $z=q^{-k}$, $k=1,2, \ldots, n$ from $(qz;q)_\infty^{-1}$ and then by taking the specialization $x=q^n$ or a contour enclosing the poles at $z=q^{-k}$, $k=1,2, \ldots n$ after the specialization $x=q^n$ for the case of the trefoil knot, we obtain the colored Jones polynomials.	\\

The examples discussed in section \ref{sec:ab} can be put in a general form.
The inverted Habiro series expression for a normalized homological block for a knot $K$ when $G_{\mathbb{C}}=SL(2, \mathbb{C})$ is given by
\begin{align}
F_K(x,q) = \sum_{k=0}^{\infty} a_{-k-1}(K;q) \frac{(-1)^k q^{\frac{1}{2}k(k+1)}}{(x;q)_{k+1} (x^{-1};q)_{k+1}}	\,	,
\label{fkinvhs}
\end{align}
while the Habiro cyclotomic expansion for a normalized Jones polynomial is given by
\begin{align}
J_K(n,q) = \sum_{k=0}^{\infty} a_{k}(K;q) (-1)^k q^{-\frac{1}{2}k(k+1)} (qx;q)_{k} (qx^{-1};q)_{k}	 \bigg|_{x=q^n}	\,	.
\label{joneshs}
\end{align}
The factor $(-1)^k q^{\frac{1}{2}k(k+1)}$ in \eqref{fkinvhs} can be obtained from the residue of $(q;q)_\infty^2 \frac{dz}{2\pi i z} \theta(-q^{\frac{1}{2}}z;q)^{-1}$ at $z=q^k$, $k=0,1, \ldots$.
For simplicity of discussion, we choose $\frac{(qzx;q)_\infty (qzx^{-1};q)_\infty}{(x;q)_{\infty} (x^{-1};q)_{\infty}}$ in the integrand to realize the factors $(x;q)_{k+1}^{-1} (x^{-1};q)_{k+1}^{-1}$ in \eqref{fkinvhs}.\footnote{The following discussion can also apply to the case where the factors $(x;q)_{k+1}^{-1} (x^{-1};q)_{k+1}^{-1}$ in \eqref{fkinvhs} are obtained from $\frac{(qzx;q)_\infty}{(z^{-1}x;q)_\infty}$ or $\frac{(qx;q)_\infty (qx^{-1};q)_\infty}{(z^{-1}x;q)_\infty (z^{-1}x^{-1};q)_\infty}$.}
In this setup, we expect that the Laurent series $a_{-k-1}(K;q)$ can be expressed in the integrand as a function of $z$ that admits a 3d $\mathcal{N}=2$ theory interpretation and recovers $a_{-k-1}(K;q)$ when $z=q^k$ as in the examples in section \ref{sec:ab}.
We denote it by $``a_{-k-1}(K;q)"(z)$.\footnote{More precisely, the monomial factors in $a_{-k-1}(K;q)$ including $q^k$ and $q^{\frac{k^2}{2}}$ would be expressed as rational functions of Jacobi theta functions in $z$ as in section \ref{sec:ab}.
In the case that the Laurent series $a_{-k-1}(K;q)$ is expressed as a sum over discrete variables or contains $q$-Pochhammer symbols, $``a_{-k-1}(K;q)"(z)$ would be obtained by replacing the sums with contour integrals over the corresponding continuous variables or by including the corresponding $q$-Pochhammer symbols with appropriate continuous variables.}
Then the homological block \eqref{fkinvhs} can be obtained from the integral, 
\begin{align}
(q;q)^2_\infty  \int \frac{dz}{2\pi i z} \frac{``a_{-k-1}(K;q)"(z)}{\theta((-q^{\frac{1}{2}})z ;q)} \frac{(qzx;q)_\infty (qzx^{-1};q)_\infty}{(x;q)_{\infty} (x^{-1};q)_{\infty}}
\label{intgen}
\end{align}
by choosing poles $z=q^k$, $k=0,1, \ldots$.

If choosing other poles $z=q^{-k}$, $k=1, 2, \ldots, n$, $``a_{-k-1}(K;q)"(z)$ becomes $a_{k-1}(K;q)$, while the residue $(q;q)_\infty^2 \frac{dz}{2\pi i z} \theta(-q^{\frac{1}{2}}z;q)^{-1}\frac{(qzx;q)_\infty (qzx^{-1};q)_\infty}{(x;q)_{\infty} (x^{-1};q)_{\infty}}$ at $z=q^{-k}$ yields the factors $(-1)^k q^{-\frac{1}{2}k(k-1)} (qx;q)_{k-1} (qx^{-1};q)_{k-1}$.
This leads to the $n$-colored Jones polynomial \eqref{joneshs} up to an overall sign after the specialization $x=q^n$.\footnote{This indicates that the Habiro coefficients and the inverted Habiro coefficients are related, in the half-index realization, by choosing different sets of poles.
It means that homological blocks can be obtained from the known expressions of the colored Jones polynomials in the form of the Habiro series.
This will be discussed further and generalized in \cite{Chung-hbN}.}

In addition, the $x$-dependence of the twisted superpotential $\widetilde{\mathcal{W}}$ from \eqref{intgen} is given by $\text{Li}_2(zx) + \text{Li}_2(zx^{-1}) + \frac{1}{2} (\log (-x))^2$, so the condition $y = \exp \partial \widetilde{\mathcal{W}} / \partial \log x$ gives rise to the abelian branch $y=1$ at $z=1$ as in \eqref{aby1}.

Therefore, in general, we expect that it is possible to obtain the half-index realization as in \eqref{intgen} for the homological blocks\footnote{We worked with the normalized version of the homological block.
The unnormalized version of the homological block is obtained by multiplying the normalized version by the unknot factor $x^{\frac{1}{2}} - x^{-\frac{1}{2}}$.
This factor can be expressed, for example, as a rational function of $q$-Pochhammer symbols, 
$x^{\frac{1}{2}} - x^{-\frac{1}{2}} = -\frac{(x^{1/2};q)_\infty (qx^{-1/2};q)_\infty (x^{-1};q)_\infty} { (q x^{1/2};q)_\infty (x^{-1/2};q)_\infty (q x^{-1};q)_\infty }$, from which we can read off the field content of the corresponding 3d $\mathcal{N}=2$ theory with boundary conditions.} expressed as inverted Habiro series \eqref{fkinvhs},\footnote{
At present, explicit inverted Habiro series expressions are known in limited cases. 
Meanwhile, an inverted Habiro series expression for $G_{\mathbb{C}}=SL(N,\mathbb{C})$ was also conjectured in \cite{Park-inverted}, so the half-index realization discussed here can be extended to $G_{\mathbb{C}}=SL(N,\mathbb{C})$. 
This generalization will be discussed in \cite{Chung-hbN}.
}
and also to obtain the colored Jones polynomial by choosing another set of poles.
We expect that this provides a 3d $\mathcal{N}=2$ theory $T[M_3]$ that captures all branches of flat connections.	\\

We discuss some other aspects of the homological block as a half-index.


\subsubsection*{$S^2 \times_q S^1$ index and homological block}

If the half-index is expressed as an integral $\mathcal{B}_\alpha = \int_{\Gamma_\alpha} \frac{dz}{2 \pi i z}\Upsilon$, it is expected that the $S^2 \times_q S^1$ partition function is given by
\begin{align}
\sum_{m} \oint \frac{ds}{2\pi i s} \large| \Upsilon \large|^2 = \sum_{\alpha} \large| \mathcal{B}_\alpha \large|^2	\,	.	\label{s2s1pf}
\end{align}
where $m$ is a magnetic flux and $\alpha$ denotes the critical points of the twisted superpotential.
The standard formula \cite{KimS, Imamura-Yokoyama, Kapustin-Willett} for the $S^2 \times_q S^1$ partition function is expressed as the left-hand side of \eqref{s2s1pf} where the integrand is given by the fusion of the holomorphic and anti-holomorphic part with appropriate assignments of parameters for each \cite{Beem-Dimofte-Pasquetti}.
In the context of the 3d-3d correspondence, $\alpha$ corresponds to flat connections up to conjugation.
In particular, for the standard calculation of the $S^2 \times_q S^1$ partition function, $\alpha$ is expected to correspond to non-abelian flat connections.
In general, taking into account all flat connections, the partition function of the complex Chern-Simons theory with integer level $\mathrm{k}=0$ or the $S^2 \times_q S^1$ partition function is weighted by the volume of the stabilizer, $\text{Stab}_\alpha$, of the flat connection $\alpha$ in $SL(2,\mathbb{C})$,
\begin{align}
\mathcal{Z}_{CS}[M_3]_{\mathrm{k}=0} = \sum_{\text{flat } \alpha} \frac{1}{\text{Vol}(\text{Stab}_\alpha)} \mathcal{Z}_\alpha	\,	.
\label{indwt}
\end{align}
The volume of the non-trivial stabilizer for abelian flat connections is infinite for a non-compact gauge group, so the contribution from the abelian flat connection vanishes in \eqref{indwt}. 
However, the contribution from non-abelian flat connections gives non-vanishing value due to triviality of the volume of the stabilizer \cite{Chung-Dimofte-Gukov-Sulkowski}.

Meanwhile, the $S^2 \times_q S^1$ index is expected to be given by the sum of homological blocks over the abelian flat connections when $M_3$ is a closed 3-manifold,
\begin{align}
\mathcal{I}(q) = \sum_{a} |\mathcal{W}_a| \hat{Z}_{a}(q) \hat{Z}_{a}(q^{-1})	
\label{indhb}
\end{align}
where $a$ denotes abelian flat connections up to conjugation and $\mathcal{W}_a$ is a stabilizer subgroup of $a$ in the Weyl group $\mathbb{Z}_2$ \cite{Gukov-Pei-Putrov-Vafa}.
It was checked when $M_3$ is a $L(p,1)$ lens space and the total space of circle bundle with degree $-p$ over the genus $g$ Riemann surface, and a similar relation is expected to hold when $M_3$ is a knot complement.
Though \eqref{indhb} is expressed as the sum of the contributions from abelian flat connections, it is expected to also contain contributions from non-abelian flat connections.
If the sum \eqref{indwt} with the weight is regularized properly, it is expected that the resulting partition function would lead to the partition function \eqref{indhb} \cite{Gukov-Pei-Putrov-Vafa}.	\\

If we take the theory $T[S^3 \backslash 4_1]$ for \eqref{hb41int} or \eqref{hb41int2} and calculate the index from the standard formula \cite{KimS, Imamura-Yokoyama, Kapustin-Willett}, the resulting index doesn't contain the abelian branch.
The index that contains the abelian branch would be obtained from \eqref{indhb} after obtaining the anti-homological block.
However, we don't have a systematic way to calculate it.\footnote{If we simply take the inversion such as $(qx;q)_\infty \rightarrow (x;q)_\infty^{-1}$ as in \cite{Beem-Dimofte-Pasquetti} for the case of the figure-eight knot, we have an integral
\begin{align}
(qx;q)_\infty (qx^{-1};q)_\infty \int \frac{dz}{2 \pi i z} \frac{(z;q)_\infty (qz^{-1};q)_\infty}{(zx;q)_\infty (zx^{-1};q)_\infty}
\end{align}
up to an overall factor.
If we take the contour to be $|z|=1$ or to enclose either set of poles arising from the $q$-Pochhammer symbols in the denominator, the resulting expression contains only the non-abelian branch but not the abelian branch, so is not appropriate for the anti-homological block.
Or we may just take the operation $q \rightarrow q^{-1}$ and $x \rightarrow \tilde{x}^{-1}$ in \eqref{hb41}, which actually gives the same expression but with $x$ replaced by $\tilde{x}^{-1}$.
However it is known that this operation is subtle \cite{CCFGH}.
If it is appropriate, which we expect to be the case for the figure-eight knot and the trefoil knots in section \ref{sec:ab}, then the generalized $S^2 \times_q S^1$ index \cite{Kapustin-Willett} that captures the abelian branch would be given by the product of \eqref{hb41} with $x$ and $\tilde{x}^{-1}$ where $x=q^{\frac{m_x}{2}}\zeta_x$ and $\tilde{x}=q^{\frac{m_x}{2}}\zeta_x^{-1}$, and similarly for the trefoil knots \eqref{f31l} and \eqref{f31r}.
}
Nevertheless, we expect that the $S^2 \times_q S^1$ index containing the abelian branch can be calculated by using \eqref{indhb} or by \eqref{indwt} with an appropriate regularization for the contribution from the abelian branch.


\subsubsection*{State-integral model incorporating abelian branch}

A state-integral model that captures the abelian flat connection was discussed in \cite{Garoufalidis-Gu-Marino-Wheeler}.
For the figure-eight knot, a factor $\tanh (\pi b^{-1} v)$ was introduced to the original state-integral
\begin{align}
\mathcal{Z}_{\text{state-int.}} \simeq \frac{\sinh(\pi b^{-1} u)}{\sinh(\pi b u)} \int_{\mathcal{C}} dv \, e^{2\pi i u (v-\frac{i}{2}b^{-1})}\tanh (\pi b^{-1} v) \frac{\Phi_{b}(-v+\frac{i}{2}b^{-1} +u)}{\Phi_{b}(v-\frac{i}{2}b^{-1} +u)}
\label{sint41}
\end{align}
where $\mathcal{C}$ is along the real axis of the $v$-plane and $v$ is the scalar component of the $U(1)$ vector multiplet.
Here, $\Phi_b(x)$ is a Fadeev quantum dilogarithm, which is
\begin{align}
\Phi_b(w) = \frac{(e^{2\pi (w+c_b)b};q)_\infty}{(e^{2\pi (w-c_b)b^{-1}};\tilde{q}^{-1})_\infty}	\,	,	\quad	c_b=\frac{i(b+b^{-1})}{2}\,	,	\quad	q=e^{2 \pi i b^2}	\,	,	\	\tilde{q}=e^{2 \pi i b^{-2}}
\end{align}
when $\text{Im} \, b^2 >0$.

In addition to the usual poles from the rational function of $\Phi_b$, which are associated with contributions from the non-abelian flat connections, there are poles $v=\pm(m+\frac{1}{2})ib$, $m=0,1, \ldots$ from the term $\tanh (\pi b^{-1} v)$.
By deforming the contour to capture poles $v=-(m+\frac{1}{2})ib$ on the negative imaginary axis of the $v$-plane, the homological block is obtained by summing the residues.
If choosing another set of poles $v=(m+\frac{1}{2})ib$ with $m=0,1, \ldots, n-1$ on the positive imaginary axis of $v$, the sum of the resulting residues gives the $n$-colored Jones polynomial when $x=q^{n}$.

By considering the relation between the $S_b^3$ partition function and the half-index, these poles $v=-(m+\frac{1}{2})ib$, $m \in \mathbb{Z}_{\geq 0}$ that gives the homological block correspond to $z=e^{2 \pi b v} = q^{-m} q^{-1/2}$ in the half-index.
We note that the poles $v=-(m+\frac{1}{2})ib$, $m \in \mathbb{Z}_{\geq 0}$ that arise from the additional factor $\tanh (\pi b^{-1} v)$ is identified with the poles $z=q^k$, $k \in \mathbb{Z}_{\geq 0}$ from $(z^{-1};q)_\infty^{-1}$ of $\theta((-q^{\frac{1}{2}})z;q)^{-1}$ discussed in section \ref{sec:ab}, if we take $z \rightarrow q^{-\frac{1}{2}}z^{-1}$ in order to match the $q$-Pochhammer symbols in \eqref{hb41int} with those of the holomorphic part of \eqref{sint41}.\footnote{
Furthermore, by taking the $\hbar \rightarrow 0$ limit, the factor $\cosh (\pi b^{-1} v)$ of $\tanh (\pi b^{-1} v)$, which is responsible for the poles, can be realized from the Jacobi theta function of the half-index.
In the $\hbar \rightarrow 0$ limit, $\theta(-z;q)^{-1}$, which is from $\theta((-q^{\frac{1}{2}})z;q)^{-1}$ with $z \rightarrow q^{-\frac{1}{2}}z^{-1}$, becomes proportional to $\cosh (\pi b^{-1} v)^{-1}$ by using properties of the $q$-Pochhammer symbol and the $(q\text{-})$Gamma function.
Thus, at the level of poles, the half-index \eqref{hb41int} provides a reason for introducing the factor $\tanh (\pi b^{-1} v)$ in \eqref{sint41} to capture the abelian flat connection.
}

Since only the factor $\tanh (\pi b^{-1} v)$ is added without its holomorphic counterpart $\tanh (\pi b v)$, the state-integral model \eqref{sint41} gives only the holomorphic part (half-index) for the abelian branch without the anti-holomorphic part (anti-half-index), while the non-abelian branch parts are given by the fusion of half and anti-half-indices with extra factors due to the inclusion of $\tanh (\pi b^{-1} v)$.
If we include $\tanh (\pi b  v)$ in the state-integral \eqref{sint41}, we obtain something else than the homological block, so it wouldn't be the desired result.
Since the squashed 3-sphere $S_b^3$ partition function is also expected to be expressed as a fusion of half- and anti-half-indices (\textit{c.f.} \eqref{s2s1pf} and \eqref{indhb}), it is not clear whether the state-integral model \eqref{sint41} would be related to the partition function in the 3d-3d correspondence.


\subsubsection*{Homological block and Stokes phenomena}

In section \ref{sec:ab}, homological blocks were obtained from the half-index \cite{Gadde-Gukov-Putrov-wall, Sugiyama-Yoshida, Dimofte-Gaiotto-Paquette}.
They can also be obtained in the context of resurgent analysis via Borel resummation of the Borel transform of the perturbative expansion around the abelian flat connections \cite{Gukov-Marino-Putrov, Gukov-Manolescu}.

For the $(P,Q)$-torus knot, the homological block is obtained from the Borel resummation \cite{Chung-resurg, Chung-seifertknot,Gukov-Manolescu},
\begin{align}
2 e^{\frac{\pi i \kappa}{2}PQu^2}\int_{\gamma} d\eta \, e^{-\frac{\kappa}{2\pi i} \frac{1}{PQ} (\eta - \pi i P Q u)^2} \frac{(e^{\frac{\eta}{P}} - e^{-\frac{\eta}{P}}) (e^{\frac{\eta}{Q}} - e^{-\frac{\eta}{Q}}) }{e^{\eta} - e^{-\eta}}		\label{Bsum-torus}
\end{align}
where $\kappa$ is the analytically continued Chern-Simons level, $\text{Re } \hbar <0$ with $\hbar = \frac{2\pi i}{\kappa}$, and $\gamma$ is an integral contour on a complex $\eta$-plane that is parallel to the imaginary axis and shifted by $+\epsilon$ along the real axis. 
Here, $u= n/\kappa$ and $n$ denotes the $n$-dimensional representation of $SU(2)$, which is analytically continued in \eqref{Bsum-torus}.
When taking $\mathrm{k} \rightarrow \mathbb{Z}_{>0}$, the integration contour $\gamma$ is deformed for convergence.
In this process, the contributions from non-abelian flat connections are recovered and, together with the integral over the deformed contour, give the Jones polynomial \cite{Gukov-Marino-Putrov, Chung-resurg}.	\\

The integral expressions for the half-index that we considered are not the type of Borel-resummed expression.
In the Borel resummation, the dependence of $q$ or $\hbar$ appears only at the Gaussian factor\footnote{It is the kernel of the Laplace transform in the standard expression of the Borel resummation.} and the convergence is controlled by the Gaussian factor, unlike the integral expression of the half-index with $q$-Pochhammer symbols.

By simply varying $q=e^\hbar$ within $|q|<1$ or approaching roots of unity from inside the unit circle $|q|<1$, unlike the case of \eqref{Bsum-torus} briefly described above, the contour enclosing the poles $z=q^k$, $k=0,1, \ldots$ doesn't need to be deformed for convergence and accordingly it doesn't necessarily cross the poles associated with the non-abelian branch and pick up their residues.
However, by varying $q$ and $x$, Stokes phenomena arise in the half-index, and it would be interesting to study the Stokes phenomena associated with the abelian flat connection as in \cite{Witten-accs} using the integral expression for the half-index discussed in section \ref{sec:ab}.
It would also be interesting to find a way at the level of the integral to map the integral expression of the half-indices with $q$-Pochhammer symbols to that of the Borel resummation as in \eqref{Bsum-torus}, from which one would be able to see directly that the homological block can capture the contributions from non-abelian flat connections.


\acknowledgments{We would like to thank T. Dimofte and S. Gukov for helpful discussions.
We would also like to thank the Korea Institute for Advanced Study (KIAS) for hospitality at various stages of this work.
This research was supported by the 2025 scientific promotion program funded by Jeju National University.
}


\begin{appendices}


\section{Additional branch in SUSY parameter space for trefoil knot}
\label{sec:appendix}

As discussed in section \ref{sec:trefoil}, in the case of the trefoil knots, an additional factor or branch, $xy+1$ or $y+x$, appears in the SUSY parameter space obtained from the half-index for a general $x \in \mathbb{C}^*$, but does not arise in the case of the colored Jones polynomial where $x=q^n$.
In this appendix, we discuss how such extra factors are related to the asymptotic behavior of the half-index and how the extra factor disappears in the case of the colored Jones polynomial of the left-handed trefoil knot.	\\

For the left-handed trefoil knot, as discussed in footnote \ref{31lnbr}, there is another solution $z=\infty$ in the critical point equation $1= \exp \partial \widetilde{\mathcal{W}}_{3_1^l} / \partial \log z$, and this solution gives rise to the branch $xy+1=0$ from $y= \exp \partial \widetilde{\mathcal{W}}_{3_1^l} / \partial \log x$.

The extra factor $xy+1$ arising from $z = \infty$ can be captured from the half-indices.
For $|q|<1$, the poles $z=q^k$ and $z=xq^k$ approach $0$, while the poles $z=q^{-k}$ approach $\infty$ as $k \rightarrow \infty$.
The half-indices are obtained by taking a contour that encloses either set of poles and extends toward 0 or $\infty$.
For the former, they are given by \eqref{31labcal} and \eqref{31lnabcal}, while for the latter it is given by \eqref{31ljones}.
As can be seen in \eqref{31lfull}, \eqref{31lnab}, and \eqref{31ljp}, the classical limits of the annihilators for the former do not contain the branch $xy+1$, whereas the classical limit of the annihilator for the latter does.
The appearance of $xy+1$ at $z = \infty$ can be understood from the fact that the contour extending towards $\infty$ captures the poles approaching $\infty$ and the factor $xy+1$ arises in the classical limit of the annihilator of the resulting half-index.

It can be made more explicit by analyzing the asymptotic behavior of the half-index.
The asymptotic behavior of the integrand $\Upsilon_{3_1^l}$ of \eqref{int31l} is given by
\begin{align}
\Upsilon_{3_1^l} \overset{\sim}{\rightarrow}  
\begin{cases} 
\exp \frac{1}{\hbar} \big[ ( \log z )^2 + \frac{1}{2} ( \log (-x) )^2 \big]	&	\text{as } |z| \rightarrow 0	\vspace{1mm}	\\
\exp \frac{1}{\hbar} \big[ -\frac{1}{2} ( \log (-x) )^2 \big]			&	\text{as } |z| \rightarrow +\infty	
\end{cases}	\,	.
\end{align}
The asymptotic behavior as $z\rightarrow \infty$ is $\exp -\frac{1}{2\hbar} \big(\log (-x) \big)^2$, and the classical limit of its annihilator is $xy+1$.
Therefore, we see that the extra branch $xy+1$ arises from the asymptotic behavior as $z \rightarrow \infty$.
This also explains that there is no additional branch for the half-index with the contour extending towards $0$, since the asymptotic behavior vanishes as $z \rightarrow 0$.

Thus, we see that analyses based on the SUSY parameter space, the choice of contour for the calculation of half-indices, and the asymptotic behavior are consistent.	\\

Furthermore, it is possible to explain why the extra factor $xy+1$ arises in \eqref{31ljones} before the specialization $x=q^n$, but doesn't arise in the colored Jones polynomial, \textit{i.e.} after $x=q^n$.
The quantity $Z_{3_1^l}(x,q):=\sum_{k=0}^{\infty} q^{k} (qx;q)_k (qx^{-1};q)_k$ from \eqref{31ljones} before taking $x=q^n$ satisfies the inhomogeneous difference equation 
\begin{align}
((q \hat{x}-1) \hat{y}+ q^2 \hat{x}^3(\hat{x}-1) ) Z_{3_1^l}(x,q) = x (q x^2-1) + (q x^2-1) \theta(-q^{\frac{1}{2}}x^{-1};q)	\,	. 
\end{align}
The term $\theta(-q^{\frac{1}{2}}x^{-1};q)$ is responsible for the asymptotic behavior $\exp -\frac{1}{2\hbar} \big(\log (-x) \big)^2$ as $z \rightarrow +\infty$.
Upon $x=q^n$, $Z_{3_1^l}(x,q)$ becomes the $n$-colored Jones polynomial $J_{3_1^l}(n,q)$.
Since $\theta(-q^{\frac{1}{2}}x^{-1};q) \big|_{x=q^n}=0$, the inhomogeneous difference equation becomes the recursion relation
\begin{align}
((q^{n+1}-1) \hat{y}+ q^{3n+2} (q^n - 1) ) J_{3_1^l}(n,q) = q^n (q^{2n+1}-1)	\,	,
\end{align} 
where $\hat{y}f(q^n)=f(q^{n+1})$ for $x=q^n$.
It is indeed the inhomogeneous recursion relation for the colored Jones polynomial $J_{3_1^l}(n,q)$ and doesn't contain the branch $xy+1$.	\\

The analysis is parallel to the case of the right-handed trefoil knot.
As discussed in section \ref{sec:trefoil}, the extra branch $y+x$ arises from the solution $z=0$ in the critical point equation.
In this case, as can be seen in \eqref{31rfull} and \eqref{31rnab}, it appears in the classical limits of the annihilators of the homological block \eqref{31rabcal} and the half-index \eqref{int31r2}, both of which are obtained from the contour enclosing the poles $z=q^k$ and $z=xq^k$, respectively, which approach 0 as $k \rightarrow \infty$.

The asymptotic behavior of the integrand $\Upsilon_{3_1^r}$ of \eqref{int31r} is given by
\begin{align}
\Upsilon_{3_1^r} \overset{\sim}{\rightarrow}  
\begin{cases} 
\exp \frac{1}{\hbar} \big[ \frac{1}{2} ( \log (-x) )^2 \big]				&	\text{as } |z| \rightarrow 0	\vspace{1mm}\\ 
\exp \frac{1}{\hbar} \big[ - ( \log z )^2 - \frac{1}{2} ( \log (-x) )^2 \big]	&	\text{as } |z| \rightarrow +\infty	\\	
\end{cases}	\,	.
\end{align}
The asymptotic behavior as $z \rightarrow 0$ is $\exp \frac{1}{2\hbar} \big(\log (-x) \big)^2$, and the classical limit of its annihilator is $y+x$.
In this case, the asymptotic behavior diverges as $z \rightarrow \infty$, so we don't take a contour that extends toward $\infty$.	\\

As a remark, in the case of the figure-eight knot, the asymptotic behavior vanishes as $z \rightarrow 0$, so it is consistent with the fact that there is no additional branch.

\end{appendices}

\bibliographystyle{JHEP}
\bibliography{ref}

\end{document}